%% file: DWI_v1_jcy3.tex
\documentclass[times,twocolumn,final]{elsarticle}

\usepackage{medima}

\input{packages}
\input{macros}

\usepackage{amsmath}
\usepackage{framed,multirow}
\usepackage{amssymb}
\usepackage{latexsym}
\usepackage{url}
\usepackage{xcolor}
\usepackage{xspace}
\usepackage{makecell}
\usepackage{tabularx}
\usepackage{algorithm}
\usepackage{pifont}
\newcommand{\cmark}{\ding{51}}%
\newcommand{\xmark}{\ding{55}}%

\newcolumntype{C}{ >{\centering\arraybackslash m{16mm}}}
\newcolumntype{M}[1]{>{\centering\arraybackslash}m{#1}}
\usepackage{hyperref}

\definecolor{newcolor}{rgb}{.8,.349,.1}

\journal{Medical Image Analysis}

\begin{document}

\verso{Chung \textit{et~al.}}

\begin{frontmatter}

\title{Simultaneous super-resolution  and motion artifact removal in diffusion-weighted MRI using unsupervised deep learning}

\author[1]{Hyungjin Chung}
\author[2]{Jaehyun Kim}
\author[2]{Jeong Hee Yoon}
\author[2]{Jeong Min Lee\corref{cor1}}
\ead{jmlshy2000@gmail.com}
\author[1]{Jong Chul Ye\corref{cor1}}
\cortext[cor1]{Corresponding authors.}
\ead{jong.ye@kaist.ac.kr}

\address[1]{Department of Bio and Brain Engineering, Korea Advanced Institute of Science and Technology (KAIST), Daejeon 34141, Republic of Korea}
\address[2]{Department of Radiology, Seoul National University College of Medicine, Republic of Korea}

\received{?}
\finalform{?}
\accepted{?}
\availableonline{?}
\communicated{?}

\begin{abstract}
Diffusion-weighted MRI  is nowadays performed routinely due to its prognostic ability, yet the quality of the scans are often unsatisfactory which can subsequently hamper the clinical utility. To overcome the limitations, here we propose a fully unsupervised quality enhancement scheme, which boosts the resolution and removes the motion artifact simultaneously. This process is done by first training the network using optimal transport driven cycleGAN with {\em stochastic degradation block} which learns to remove aliasing artifacts and enhance the resolution, then using the trained network in the test stage by utilizing bootstrap subsampling and aggregation for motion artifact suppression. We further show that we can control the trade-off between the amount of artifact correction and resolution by controlling the bootstrap subsampling ratio at the inference stage. To the best of our knowledge, the proposed method is the first to tackle super-resolution and motion artifact correction simultaneously in the context of MRI using unsupervised learning. We demonstrate the efficiency of our method by applying it to both quantitative evaluation using simulation study, and to in vivo diffusion-weighted MR scans, which shows that our method is superior to the current state-of-the-art methods. The proposed method is flexible in that it can be applied to various quality enhancement schemes in other types of MR scans, and also directly to the quality enhancement of apparent diffusion coefficient maps.
\end{abstract}

\begin{keyword}
\MSC[2021] 92C55 \sep 68U10\sep 34A55
\KWD \\
Diffusion weighted MRI \\
Unsupervised learning \\
Super resolution \\
Artifact removal
\end{keyword}

\end{frontmatter}


\section{Introduction}
Over the years, interest in Diffusion Weighted MRI (DW-MRI) has continuously increased thanks to its versatile ability in diverse clinical applications~\citep{tamada2017prostate, razek2018assessment, razek2019multi}. {DW-MRI quantifies the degree of Brownian motion of protons and can detect diffusion restriction in areas with high cellularity (i.e. malignancies) and cell membrane density. Detecting changes in diffusion at the cellular level means that DW-MRI can be used not only to detect and characterize lesions, but also to monitor and predict therapeutic responses.~\citep{taouli2010diffusion, pereira2019diffusion}} Another great advantage of DW-MRI comes from its ability to {\em quantify} the amount of diffusion via calculating the apparent diffusion coefficient (ADC) map, which further enhances the prognostic power.

In order to achieve DW-MRI, diffusion sensitizing bi-polar gradient is used to capture the small molecular movement. While sensitizing for microscopic movement of molecules, macroscopic patient motion can also be captured in the image, which introduces one of the greatest technical difficulties in DW-MRI. Motion artifacts arising from bulk patient motion can be reduced by cardiac gating, pulse triggering, or designing a specific pulse sequence for motion artifact removal~\citep{bammer2003basic, chilla2015diffusion}. However, these approaches cannot completely remove the artifacts, leaving much room for improvement.

\begin{figure}[!t]
    \centering\includegraphics[width=8.5cm]{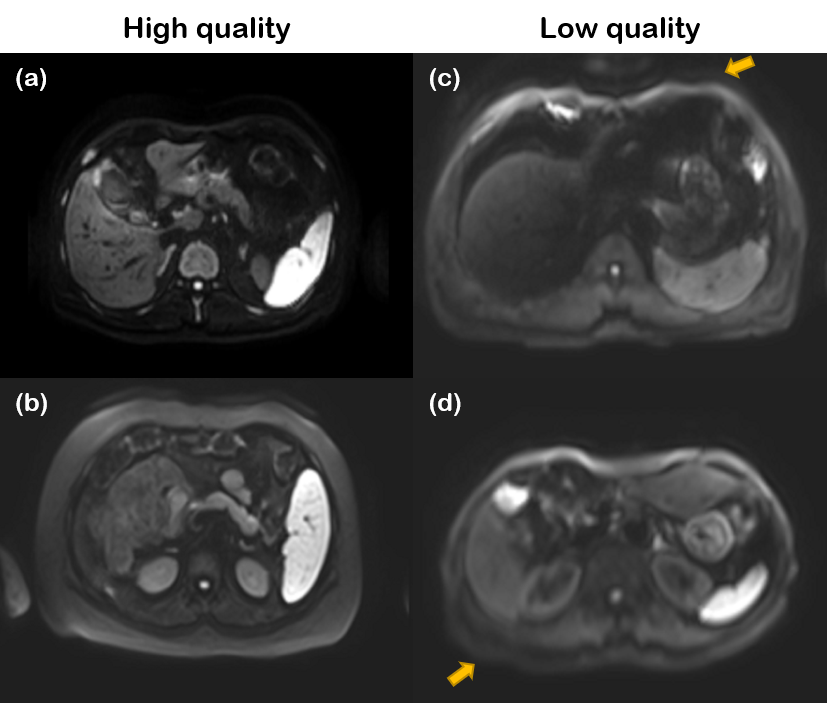}
    \caption{Illustration of variability in image quality existent in diffusion-weighted MRI. (a-b) High quality images with high resolution and no artifacts. (c-d) Low quality images with low resolution and motion artifacts. Yellow arrows indicate areas where motion artifacts are observed.}
	\label{fig:goodbad}
\end{figure}

Yet another problem in DW-MRI comes from its inherently low resolution and signal-to-noise (SNR) ratio owing to the typical usage of fast imaging schemes such as echo planar imaging (EPI)~\citep{baliyan2016diffusion} with small number of excitations (NEX). The low spatial resolution provides further challenges in DW-MRI, making it hard to visualize small lesions of interest, or exactly quantify the diffusion property. A typical example of (relatively) high quality and low quality DW-MRI images are depicted in Fig.~\ref{fig:goodbad}. Although the four different images in Fig.~\ref{fig:goodbad} were scanned with the same parameters in the same scanner, compared to images in (a-b), the images in (c-d) have lower resolution, and motion artifacts make the image quality worse.

Over the years, researchers have sought to overcome the challenges in DW-MRI with various methods. Super-resolution (SR) algorithms in DW-MRI were proposed to increase the spatial resolution of DW-MRI initially via simple interpolation based methods. Later on, model-based optimization methods~\citep{scherrer2011super} were adopted. Based on the theory of compressed sensing (CS), regularized optimization methods, most of which utilizes variants of total variation (TV) regularization, were proposed in the context of diffusion magnetic resonance imaging~\citep{ning2016joint, teh2020improved}. Other advanced methods such as the Metropolis-Hastings algorithm was proposed in~\citep{yap2013generative}. In parallel, motion artifact correction methods in DW-MRI were developed. \citep{ozaki2013motion} proposed to design a velocity-compensated diffusion gradients in order to compensate for cardiac motion. Independent of the weighting scheme, CS-based MRI motion artifact correction methods~\citep{vasanawala2010improved, yang2013sparse, jin2017mri} were also proposed. 

The downside of model-based optimization methods and CS-based methods come from their high computational complexity and cumbersome hyper-parameter tuning. Moreover, these algorithms typically require raw $k$-space data, which are often hard to collect. Moreover, most of the algorithms developed aim at a specific task, either acceleration or super-resolution, which cannot handle arbitrary degradation in the image.

Lately, deep learning based methods for DW-MRI super-resolution \citep{albay2018diffusion, tanno2017bayesian, fan2020generative} were developed, thanks to its ability to learn from data distribution. Motion artifact correction methods based on deep learning were also extensively studied~\citep{duffy2018retrospective, pawar2018motion}, but most of these studies were restricted to the supervised learning situation, where the network learns from matched target data. 

This greatly shrinks the applicability of these algorithms, since there are many situations in DW-MRI where such supervised training is unavailable. For instance, matching pairs of high-resolution (HR) and low-resolution (LR) DW-MRI scans are unobtainable. Moreover, exactly matching motion-free images are impossible to obtain. To this end, unsupervised learning-based methods such as Cycle-MedGAN~\citep{armanious2019unsupervised, armanious2020unsupervised} and CycleGAN with bootstrap aggregation~\citep{oh2020unsupervised} have been proposed. However, to the best of our knowledge, a method which simultaneously corrects for motion artifacts and performs super-resolution at once, has not been reported in literature.

To address the problem in an unsupervised fashion to be able to directly apply in real-world situation, and to solve the problem of super-resolution and artifact correction simultaneously, here we propose an unsupervised learning to solve it all. Our idea is based on the investigation of using bootstrap aggregation for probabilistic $k$-space correction~\citep{oh2020unsupervised}, which is further extended in this study to also account for super-resolution, and adapted to the quality enhancement of DW-MRI. Specifically, in the training of MR reconstruction network for $k$-space correction, we  add a {\em stochastic degradation} block as a forward mapping, which models the degradation process of the resolution in the image acquisition. By this simple extension, the $k$-space correction network learns not only to correct for under-sampled $k$-space, but also to enhance the resolution. Once the network is trained, bootstrap sub-sampling and aggregation is performed at the inference stage, where $k$-space phase encoding lines are randomly sub-sampled, and the corrected images are aggregated to form a final high-quality reconstructed image.

The applicability of our novel deep learning framework was explored using both simulated data, and in vivo liver DW-MRI scans. Results confirm that the proposed method can effectively remove motion artifacts present in DW-MRI and increase the resolution/SNR altogether. Interestingly, the same reconstruction steps can also be directly applied to enhance the quality of the ADC maps, showing the versatility of the proposed method.

\section{Theory}

\begin{figure*}[!t]
    \centering\includegraphics[width=15cm]{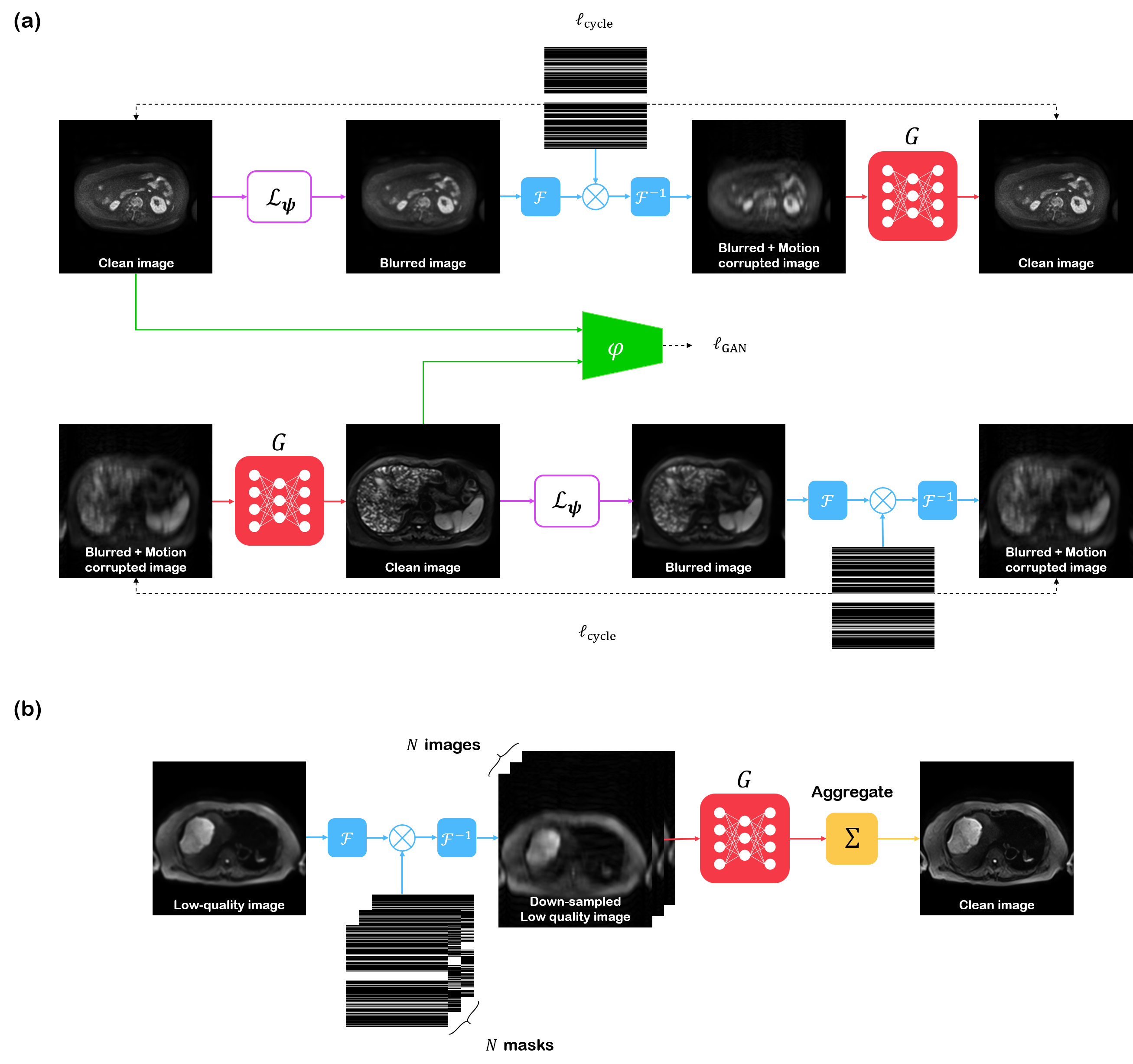}
    \caption{Pictorial description of the overall method. (a) Unsupervised training scheme for the proposed method. At each step, $\boldsymbol{\psi}$ is sampled from $P_{\boldsymbol{\psi}}$, which blurs the image and adds noise. Then, $k$-space under-sampling is performed and the network $G$ is trained to remove the blurring $+$ motion artifacts. (b) Boosting aggregation for inference. Following~\citep{oh2020unsupervised}, the degraded images are randomly sub-sampled in the $k$-space and reconstructed with the trained network $G$, after which follows averaging of the images.}
	\label{fig:proposed_network}
\end{figure*}

Consider the following forward measurement model:
\begin{equation}
    \yb = \Fc {\xb_L},
\label{eq:measurement}
\end{equation}
where $\yb$ is the  $k$-space data, $\Fc$ corresponds to Fourier transform, and ${\xb_L}$ is the low-quality image in which the resolution is limited by the system, formally defined as:
\begin{equation}
\begin{aligned}
    \xb_L = \Lc_{\boldsymbol{\psi}}\xb, \,
\quad \mbox{where}   \quad \boldsymbol{\psi} \sim P_{\boldsymbol{\psi}}.
\label{eq:degradation}
\end{aligned}
\end{equation}
Here, $\xb$ is the latent high-quality image that we would like to estimate, 
$\Lc_{\boldsymbol{\psi}}$ is the stochastic degradation process parametrized by a random sample $\boldsymbol{\psi}$ from a probability
distribution $P_{\boldsymbol{\psi}}$, 
which generates the low-quality image ${\xb_L}$. Moreover, it is often the case the transient patient motion introduces $k$-space
outliers along the phase encoding direction~\citep{oh2020unsupervised}:
\begin{equation}
    \yb_c(\kappa_x, \kappa_y) = \left(\Cc\yb\right)(\kappa_x, \kappa_y) :=
    \begin{cases}
    \yb(\kappa_x, \kappa_y) e^{-\iota\Phi(\kappa_y)},\, \kappa_y \in \mathbb{I} \\
    \yb(\kappa_x, \kappa_y),\, \text{otherwise}.
    \end{cases}
\label{eq:motion}
\end{equation}
In Eq.~\eqref{eq:motion}, $\yb_c$ is the motion-corrupted $k$-space measurement, $\yb$ is the clean data in \eqref{eq:measurement}, $\kappa_x$ is the frequency-encoding direction, $\kappa_y$ is the phase-encoding direction, and $\iota = \sqrt{-1}$. Here, note that  motion artifacts are induced by the phase shift along the phase-encoding direction $\Phi(\kappa_y),\, \kappa_y \in \mathbb{I}$, where $\mathbb{I}$ contains the indices of errors.  Eqs.~\eqref{eq:measurement},~\eqref{eq:degradation} and \eqref{eq:motion} tells us that
the forward model for the measurement is given by
\begin{align}
\yb_c=\Cc \Fc \Lc_\psib \xb
\end{align}
Hence, the correction of $k$-space outliers, together with the inversion of the degradation process by blur operation $\Lc_\psib$ is necessary in order to recover  clean images at the desired resolution. 

\subsection{Bootstrap Aggregation for Artifact Correction and Super Resolution}

Recently in our companion paper~\citep{oh2020unsupervised}, we demonstrated that the $k$-space outliers can be corrected by a bootstrap aggregation approach in the context of MRI reconstruction~\citep{cha2020geometric}, which can be formulated as follows:
\begin{equation}
    \tilde{\xb} = \sum_{n=1}^N w_n G\left(\Fc^{-1}\Tc_n \yb\right),
\label{eq:bootstrap_agg}
\end{equation}
where $\Tc_n$ is the random down-sampling operator which probabilistically masks out phase-encoding lines, $\yb$ is the $k$-space measurement, $w_n$ is the weight applied to the $n^{\text{th}}$ reconstruction,  $\Fc^{-1}$ refers to the inverse Fourier transform, and $G$ is the trained neural network which can reconstruct high quality images from the aliased input image due to under-sampled $k$-space measurement. In other words, the measured $k$-space is randomly sub-sampled $N$ different ways, reconstructed via the reconstruction network $G$, and subsequently aggregated to form the final image.

The reason why the approach in \eqref{eq:bootstrap_agg} is efficient for motion artifact correction comes from the following observation
\begin{equation}\label{eq:approx}
    \Tc_n \yb \simeq \Tc_n \yb_c = \Tc_n\Cc\yb,
\end{equation}
for certain $n$, where the operator $\Cc$ is defined in \eqref{eq:motion}. In another words, when the motion corrupted $k$-space outliers in the phase-encoding lines are removed, the sub-sampled image becomes similar to the sub-sampled image from the clean image. Hence, for certain sampling pattern $\Tc_n$, the motion-corrupted $k$-space outlier can be effectively removed. Subsequently, the bootstrap aggreagtion scheme introduced in Eq.~\eqref{eq:bootstrap_agg} corrects for the missing $k$-space components to form a clean image, since the network $G$ was trained to reconstruct the clean images. 

\begin{figure}[!t]
    \centering\includegraphics[width=8cm]{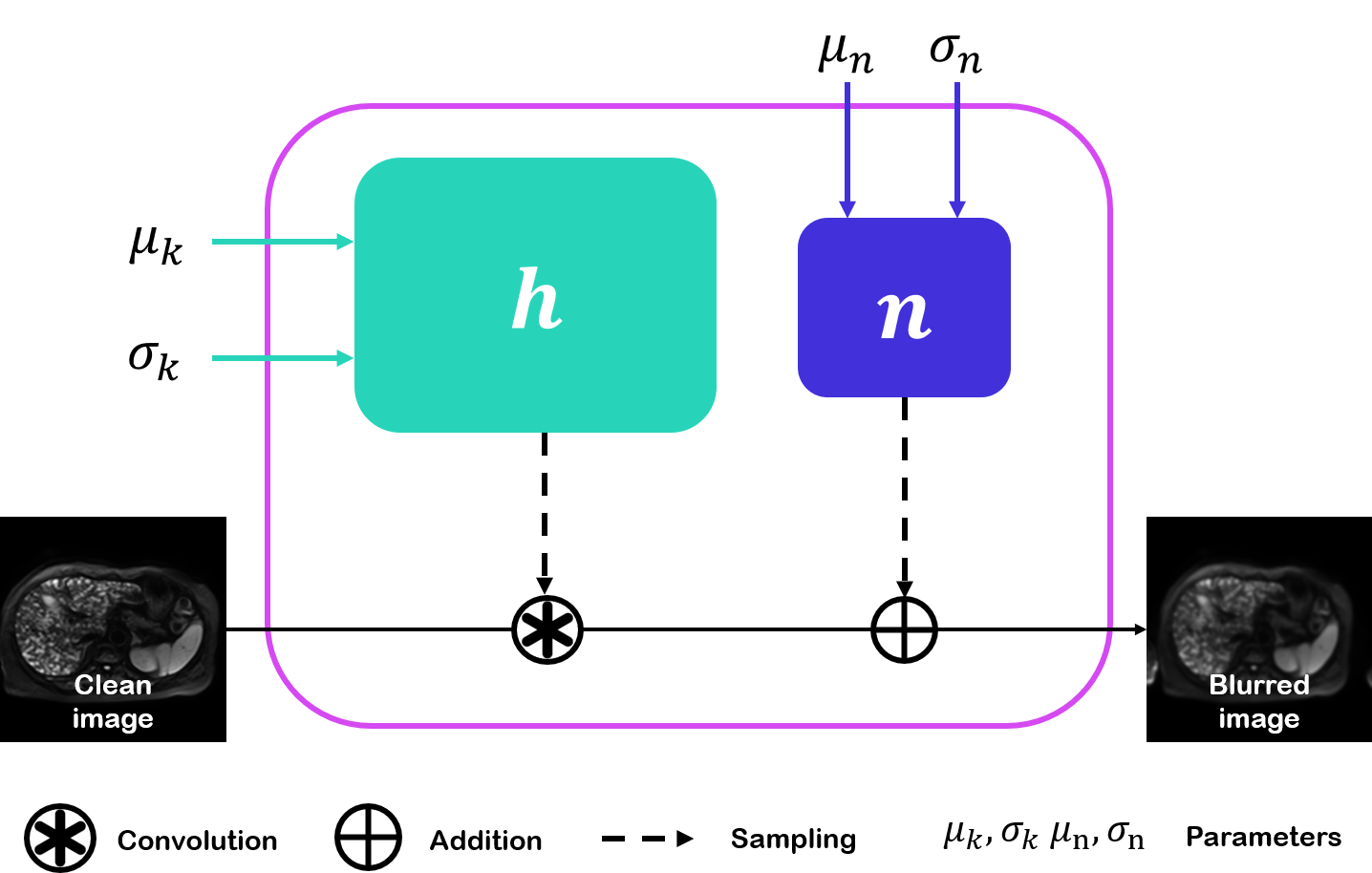}
    \caption{Pictorial representation of the degradation block $\Lc_{\boldsymbol{\psi}}$. $\boldsymbol{\psi} = \{\mu_k, \sigma_k, \mu_n, \sigma_n\}$ are the parameters defining $\Lc_{\boldsymbol{\psi}}$. Subsequently, the convolution kernel and noise are stochastically sampled.}
	\label{fig:degradation_block}
\end{figure}

Unfortunately, this is not enough in the case where the image to be reconstructed through bootstrap aggregation is also degraded by \eqref{eq:degradation}, and hence has lower resolution. To account for this additional caveat, we extend the training framework to incorporate the image degradation process in \eqref{eq:measurement},\eqref{eq:degradation}, and \eqref{eq:motion}. Specifically, we model the degradation process $\Lc_{\boldsymbol{\psi}}$ as follows:
\begin{align}
    \Lc_{\boldsymbol{\psi}}\xb := \hb_{(\mu_k, \sigma_k^2)} \ast \xb + \nb,
\label{eq:degradation_spec}
\end{align}
where $\hb_{(\mu_k, \sigma_k^2)}$ represents the Gaussian blur kernel parameterized with $\mu_k, \sigma_k$, and the noise $\nb$ is modeled with white Gaussian noise $\nb \sim \Nc(\mu_n, \sigma_n^2)$, so that
$\boldsymbol{\psi} = \{ \mu_k, \sigma_k, \mu_n, \sigma_n \}$ is the set containing all the parameters of degradation.
Here, we further assume that $\mu_k = 0$ is fixed, and sample $\sigma_k \sim \text{Beta}(a, b)$. Furthermore, we also assume  $\mu_n = 0$ so that the inherent noise do not get emphasized while the super-resolution process. {Pictorially, $\Lc_{\boldsymbol{\psi}}\xb$ can be described as Fig.~\ref{fig:degradation_block}.} Notably, although this might not capture the true degradation process, it closely resembles the truth, as we empirically show in the Results section. Moreover, when the forward model is exactly known, our proposed method can better approximate the desired clean image, which we show in the simulation study.

All in all,  the resulting forward model for the Fourier inversion from the random $k$-space subsampling is given by
\begin{align}\label{eq:z}
\zb_n  &:=  \Fc^{-1}\Tc_n\yb_c =  \Fc^{-1}\Tc_n \Cc \Fc \Lc_\psib \xb \notag\\
&\simeq  \Fc^{-1} \Tc_n \Fc \Lc_\psib \xb
\end{align}
where $\Lc_{\boldsymbol{\psi}}$ is defined in \eqref{eq:degradation_spec}, and the last approximation is due to \eqref{eq:approx}.
Therefore, we need to train the reconstruction network such that the network learns the inverse mapping with respect to  the forward operation $\Tc_n \Fc \Lc_\psib$. For training, we utilize physics-informed OT-cycleGAN, which was shown to be especially effective for MRI reconstruction~\citep{oh2020unpaired, chung2021two, cha2020unpaired}. 
  
Specifically, let $\Xc$ and $\Zc$ denote the domains for the motion artifact free high resolution images and the aliased images from the forward operation in \eqref{eq:z}, respectively. We further assume that  $\Xc$ and $\Zc$ are associated with probability measure $\mu$ and $\nu$, respectively. We define the transport cost between the two probability spaces $(\Xc,\mu)$ and $(\Zc,\nu)$:
\begin{align}
c(\xb,\zb; G)= \|\zb - \Fc^{-1} \Tc_n \Fc \Lc_\psib \xb\| + \|\xb - G(\zb)\|
\end{align}
where $G:\Zc\mapsto \Xc$ is the neural network generator to perform inverse operator. Then, the corresponding optimal transport problem is given by
\begin{align}\label{eq:unsupervised}
\inf\limits_{\pi \in \Pi(\mu,\nu)}\int_{\Xc\times \Zc}c(\xb,\zb;G) d\pi(\xb,\zb) 
\end{align}
where $\Pi(\mu,\nu)$ denotes the set of the joint distribution whose marginals are $\mu$ and $\nu$. The geometric meaning of \eqref{eq:unsupervised} was discussed in detail in \citep{oh2020unpaired, chung2021two, cha2020unpaired}, which aims to minimize the statistical distances between  transported  and the empirical measures in $\Xc$ and $\Zc$ simultaneously.

Furthermore,  the primal formulation of the unsupervised learning in \eqref{eq:unsupervised}  can be represented by a dual formulation:
\begin{eqnarray}\label{eq:OTcycleGAN}
\min_{G}\max_{\varphi}\ell_{cycleGAN}(G,\varphi)
\end{eqnarray}
where 
\begin{eqnarray}
\ell_{cycleGAN}(G,\varphi):=  \lambda \ell_{cycle}(G) +\ell_{Disc}(G,\varphi).
\end{eqnarray}
Here, $\lambda>0$ is the hyper-parameter, and  the cycle-consistency term is given by
\begin{align}\label{eq:cycleloss} 
\ell_{cycle}(G)  =& \int_{\Xc} \|\xb- G(\Fc^{-1} \Tc_n \Fc \Lc_\psib(\xb)) \|  d\mu(\xb) \\
&+\int_{\Zc} \|\zb-\Fc^{-1} \Tc_n \Fc \Lc_\psib G(\zb)\|   d\nu(\zb), \notag
\end{align}
whereas  the second term is the discriminator term:
\begin{align}
&\ell_{Disc}(G,\varphi)  \label{eq:Disc} \\
=&\max_{\varphi}\int_\Xc \varphi(\xb)  d\mu(\xb) - \int_\Zc \varphi(G(\zb))d\nu(\zb),
\end{align}
where $\varphi$ is a 1-Lipschitz function. Note that only a single generator/discriminator pair is used for distribution mapping, since the mapping from $\Xc$ to $\Zc$ is replaced by a deterministic forward operator in \eqref{eq:z}. As discussed in \citep{oh2020unpaired, chung2021two, cha2020unpaired}, the discriminator term in \eqref{eq:Disc} can be replaced by LS-GAN~\citep{mao2017least} loss.

The resulting OT-cycleGAN  framework is illustrated in Fig.~\ref{fig:proposed_network}(a), where in the upper part of the cycle, clean image is blurred via the stochastic degradation operation $\Lc_{\boldsymbol{\psi}}$, Fourier transformed to $k$-space for sub-sampling, then transformed back to image space, after which follows the generator $G$, which learns to invert the forward mapping. The lower branch starts from the unpaired artifact image, and does the opposite. During the learning process, the discriminator $\varphi$ is used for learning the distribution of the clean image.

At test stage, as shown in Fig.~\ref{fig:proposed_network}(b), the low-quality image is Fourier transformed to $k$-space, and it is randomly subsampled using $N$ different masks, creating $N$ images containing aliasing artifacts. These images are reconstructed in parallel with the trained network $G$, which is aggregated by Eq.~\eqref{eq:bootstrap_agg}. For simplicity, for all the following experiments, we set $w_n = 1/N, \, n = \{1, \dots, N\}, \, N = 15$. 

Note that the reconstruction need not be started from raw $k$-space. DICOM images can be directly utilized, and hence our method can be applied in a wide variety of  clinical situations, where the $k$-space raw data is not available. Another great advantage is that the same neural network $G$ can be used with different sub-sampling factors at the test stage, adapting to user needs. When the sub-sampling factor is set to low values (e.g. 1.5, 2), reconstructed images are kept sharp without the presence of aliasing artifacts. When the factor is set to high values (e.g. 3, 4), bulk motion is best removed, albeit minor aliasing artifacts are introduced. For all the reported results, we set $R = 1.5$. The results and consequences will further be discussed in Section~\ref{sec:sampling_factor}.

\section{Methods}
\subsection{Training Dataset}
\label{sec:training_dataset}

To demonstrate the efficacy of our method, we initially performed a simulation study using human connectome project (HCP) dataset, containing magnitude only MR brain images. Specific parameters are shown in Table~\ref{table:HCP}. Out of 190 patient volume scans, 150 volumes (3000 slices) were used for training, and 40 volumes (800 slices) for testing. To simulate the situation where the images are contaminated with motion artifacts along with blurring, we specify the parameters of the degradation to $\mu_k = 0, \sigma_k \sim \text{Beta}(a = 2.0, b = 2.0)$, and $\sigma_n =0.01$ Procedure for motion artifact generation can be found in Section~\ref{sec:motion_artifact}.

Moreover, another simulation study using the liver DW-MRI data was performed. Specifically, out of 100 high quality DWI patient scans (4831 slices), 80 patients were selected as training set and 20 patients were selected as test set. Simulated degradation $\Lc_{\boldsymbol{\psi}}$ was set identical to the HCP simulation study, and the motion artifact generation process is elaborated in \ref{sec:motion_artifact}.

\begin{table}[!hbt]
\begin{center}
	\resizebox{0.35\textwidth}{!}{
    \begin{tabular}{l|c}
        \hline
    Parameter     & Specification\\
    \hline\hline
    Scanner     & Siemens 3T scanner \\
    Echo train duration     & 1105 \\
    Time of Repetition(TR) [ms] & 3200 \\
    Time of Echo (TE) [ms] & 565 \\
    Matrix size & 320 $\times$ 320 \\
    Resolution [mm$^3$] & 0.7 $\times$ 0.7 $\times$ 0.7 \\
    \hline
    \end{tabular}}
\caption{Specifications for the 3D HCP dataset.}
\label{table:HCP}
\end{center}
\end{table}

For in vivo study, we use 2D diffusion weighted MRI (DW-MRI) of liver acquired through single shot echo planar imaging (EPI) with specific parameters as specified in Table~\ref{table:DWI_spec}. All scans were taken as breath-hold MR scans. Moreover, scans were acquired via varying $b = 0, 30, 500, 800$ to quantify the ADC map.

\begin{table}[!hbt]
\begin{center}
	\resizebox{0.35\textwidth}{!}{
    \begin{tabular}{l|c}
            \hline
    Parameter     & Specification\\
    \hline
    \hline
    Scanner     & Siemens Skyra 3T scanner \\
    Time of Repetition(TR) [ms] & 1800 \\
    Time of Echo (TE) [ms] & 56 \\
    Echo Train Length (ETL) & 37 \\
    Matrix size & 232 $\times$ 180 \\
    Num. Excitation (NEX) & 2 \\
    Resolution [mm$^2$] & 2.0 $\times$ 2.0 \\
    Slice Thickness [mm] & 6.0 \\
    \hline
    \end{tabular}
    }
\caption{Specifications for the 2D In Vivo DWI acquisition.}
\label{table:DWI_spec}
\end{center}
\end{table}

Among 201 DWI patient scans, 100 (4831 slices) were manually classified as {\em bad quality}, and 101 as {\em good quality} (3815 slices) by an attending radiologist. Consequently, the high quality images were used as training in Fig.~\ref{fig:proposed_network}, and the low quality images were used for inference.
Notably, all the patient scans share the same scan parameters as in Table~\ref{table:DWI_spec}, and thus there are no clear distinctions between the images in two different distributions. In other words, the high quality images do not have much higher image resolution than the low quality images, which could hamper the training since target distribution is not well-defined. Nevertheless, our proposed method is still able to enhance the quality of DW-MRI images in a fully unsupervised fashion, as will be shown in the results section.

\subsection{Network Architecture}
\label{sec:net_arch}

\begin{figure}[!hbt]
    \centering\includegraphics[width=8.5cm]{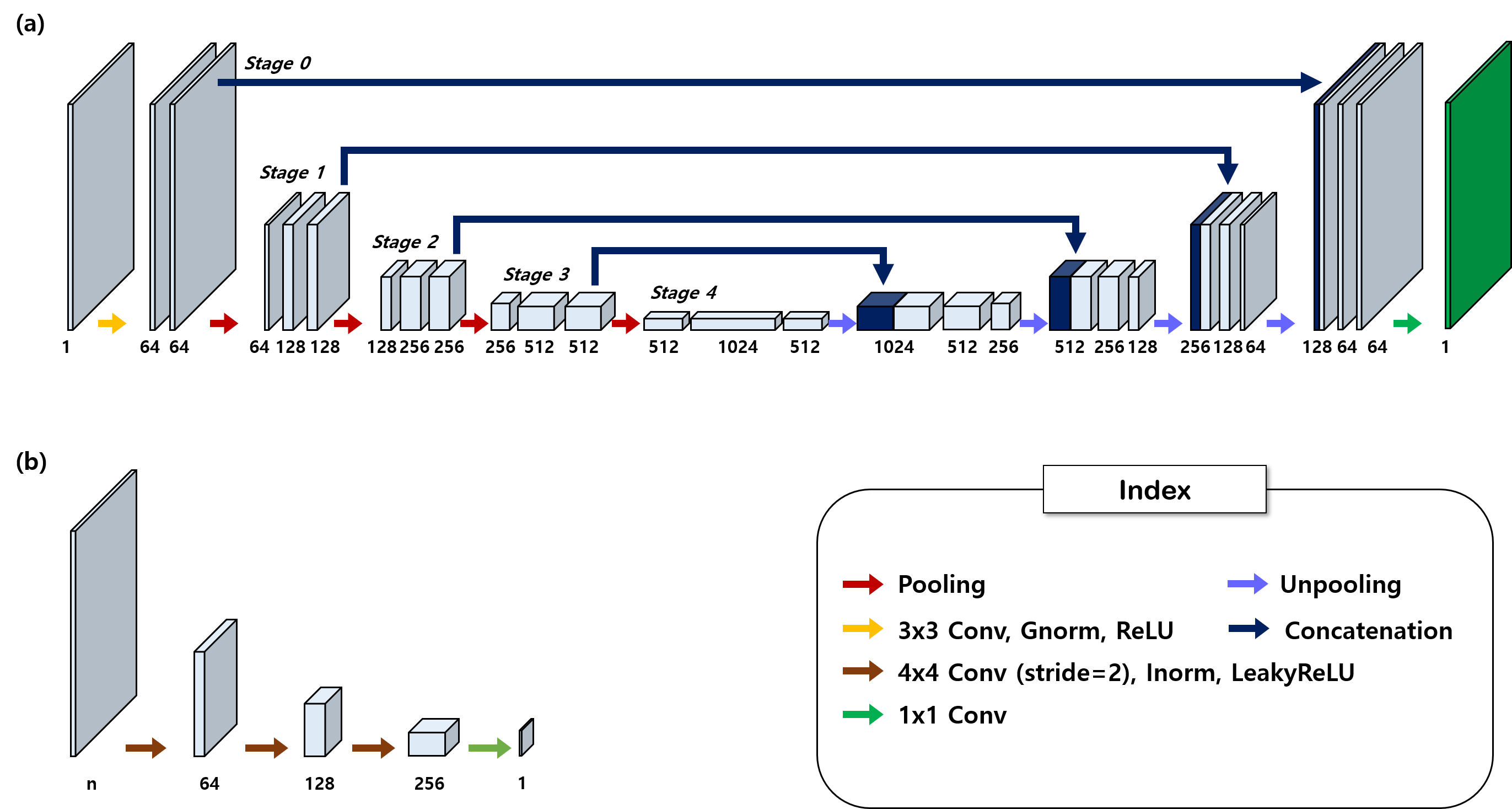}
    \caption{Network architecture used for training and bootstrap aggregation. (a) Generator architecture adopted from U-Net~\citep{ronneberger2015u}, (b) PatchGAN Discriminator architecture adopted from~\citep{isola2017image}.}
	\label{fig:net_arch}
\end{figure}

\begin{figure*}[!hbt]
    \centering\includegraphics[width=18cm]{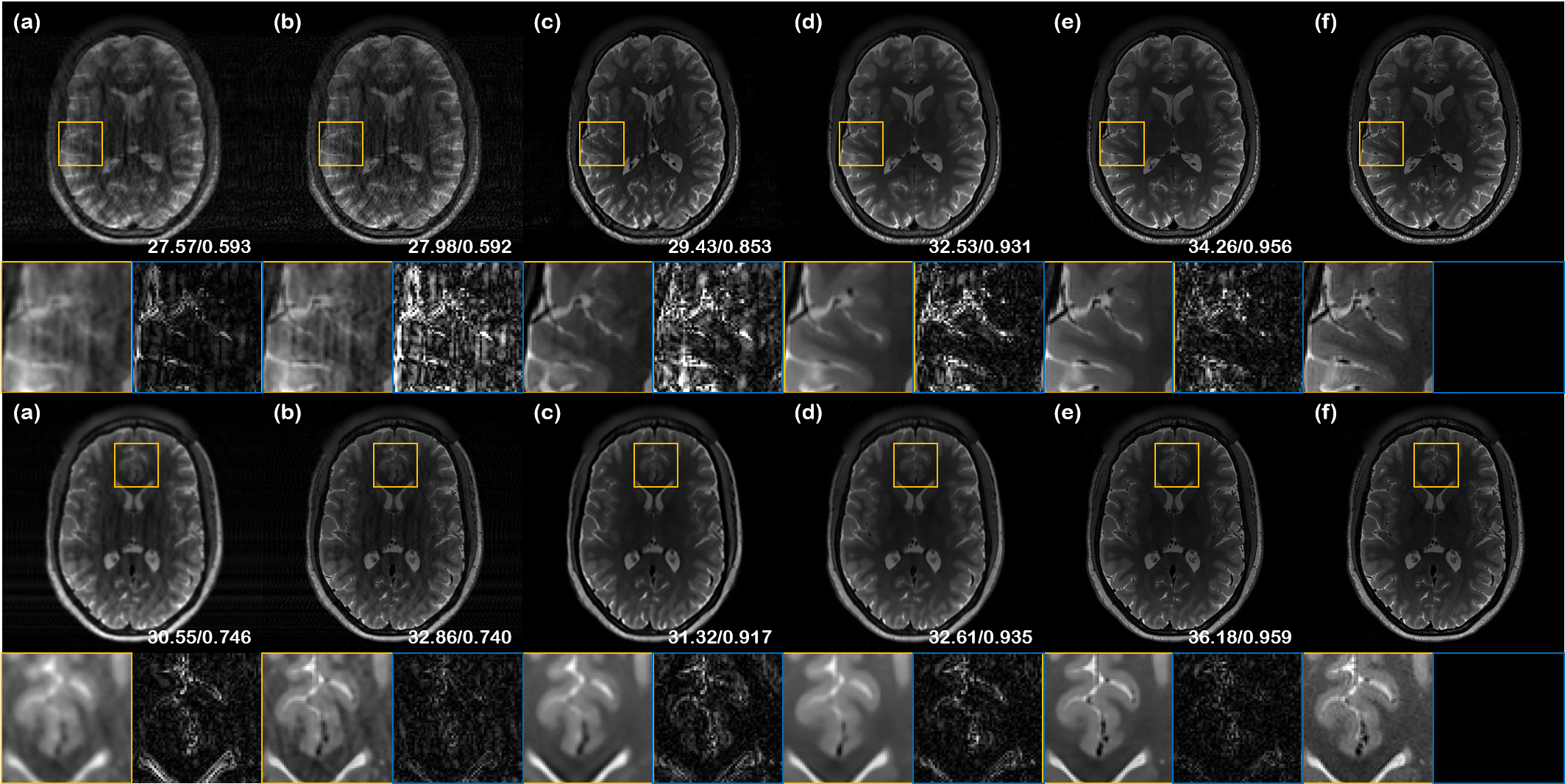}
    \caption{Reconstruction results of HCP data simulation study: (a) input images with artifact, and reconstruction results  with (b) ESRGAN~\citep{wang2018esrgan}, (c) cycleGAN~\citep{zhu2017unpaired}, (d) Oh et al.~\citep{oh2020unsupervised}, and (e) the proposed method. (f) shows the label images. Images in the yellow box show the magnified version of the specific patch, and the images in the blue blox represent the difference. Numbers in white indicate PSNR, and SSIM values, respectively.}
	\label{fig:results_HCP}
\end{figure*}

\begin{figure*}[!hbt]
    \centering\includegraphics[width=18cm]{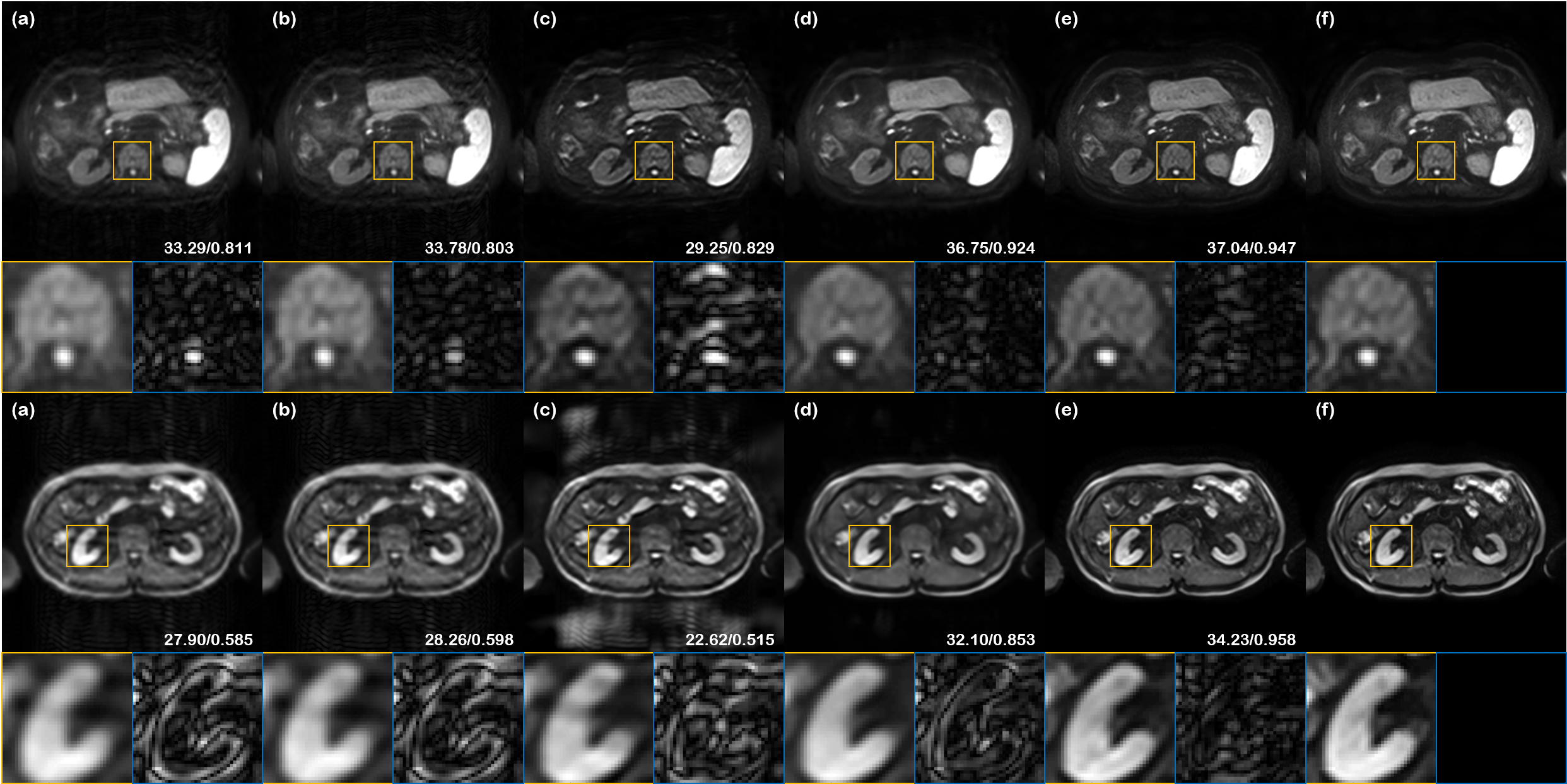}
    \caption{Reconstruction results DW-MRI simulation study: (a)  input images with artifact, and reconstruction results  with (b)ESRGAN~\citep{wang2018esrgan}, (c) cycleGAN~\citep{zhu2017unpaired}, (d) Oh et al.~\citep{oh2020unsupervised}, and (e) proposed method. (f) shows the label images. Images in the yellow box show the magnified version of the specific patch, and the images in the blue blox represent the difference. Numbers in white indicate PSNR, and SSIM values, respectively.}
	\label{fig:results_DWI_sim}
\end{figure*}

\begin{figure*}[!hbt]
    \centering\includegraphics[width=18cm]{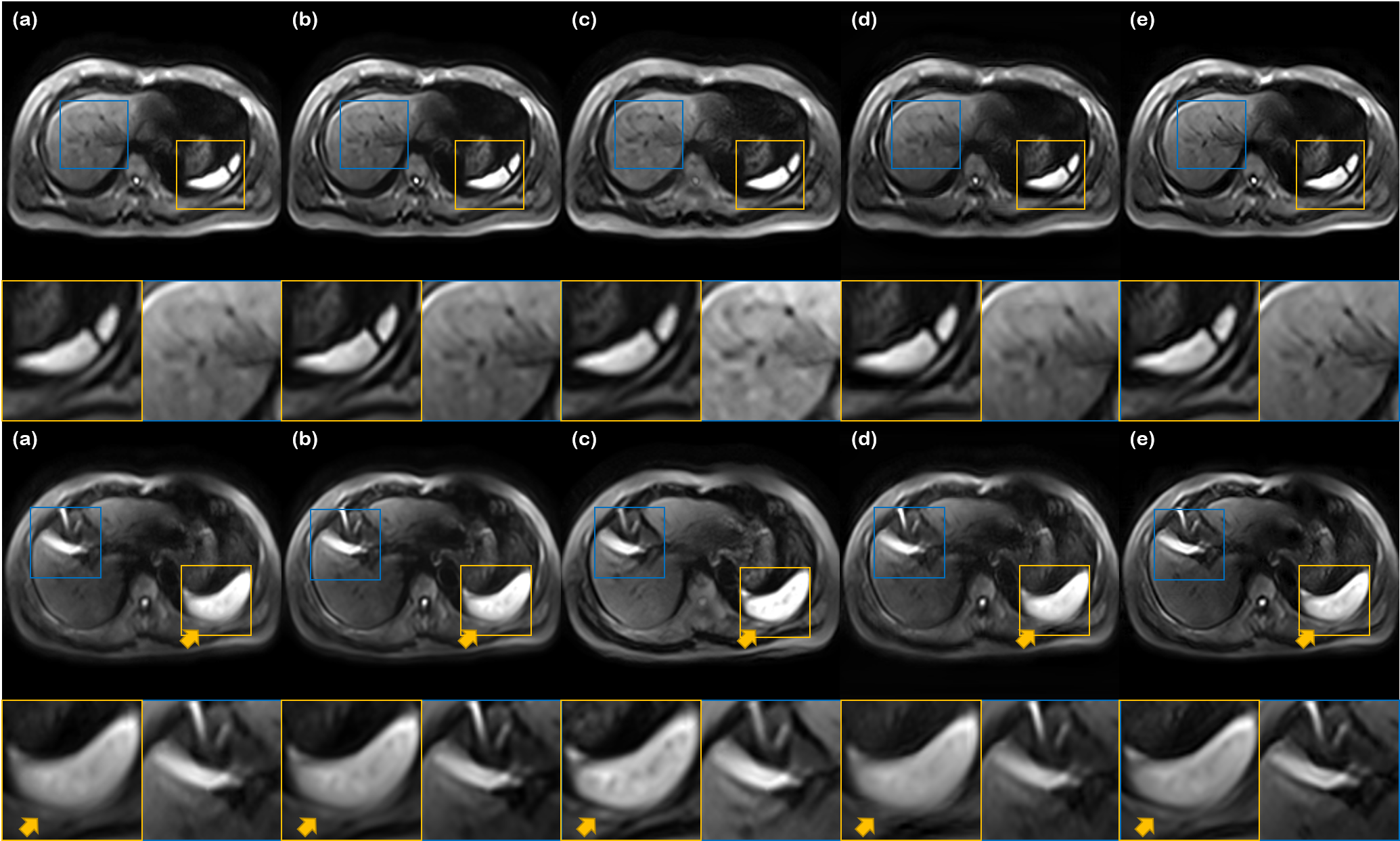}
    \caption{Reconstruction results of in vivo DW-MRI acquisition:  (a)  input images with artifact, and reconstruction results  with (b) ESRGAN~\citep{wang2018esrgan}, (c) cycleGAN~\citep{zhu2017unpaired}, (d) Oh et al.~\citep{oh2020unsupervised}, and (e) the proposed method. Images in the yellow, and blue blox show the magnified regions of each image. Yellow arrow indicates the region where motion artifact makes the boundary blurry, while the proposed method removes the artifact successfully.}
	\label{fig:results_DWI}
\end{figure*}

Network architecture for the generator $G$ and discriminator $\varphi$ used in Fig.~\ref{fig:proposed_network} (a),(b) are illustrated in Fig.~\ref{fig:net_arch}. The architecture of CNN generator depicted in Fig.~\ref{fig:net_arch}(a) was adopted from U-Net~\citep{ronneberger2015u}, with 4 stages of pooling/unpooling, the number of filter channels initially set to 64 at stage 0, which doubles at every stage, reaching 1024 at stage 4. At each stage, blocks are made up of 3 $\times$ 3 convolution, group normalization with group $= 8$, and ReLU activation. The last feature activation of the encoder stage are concatenated to the decoder layer of the corresponding stage, and at the last layer, 1 $\times$ 1 convolution is used.

For the discriminator, 2D PatchGAN architecture was adopted from~\citep{isola2017image}, as shown in Fig.~\ref{fig:net_arch}(b). Specifically, each block consists of 4$\times$4 convolution with stride$= 2$, Instance normalization, and LeakyReLU activation. For additional stability, we also used spectral normalization at every convolution block. Last layer of the network consists of 1$\times$1 convolution to get the final output.

\subsection{Training Details}

For training, the hyper-parameter for the loss function was set to $\lambda = 0.001$, and the networks were trained for 150 epochs with Adam optimizer using the default parameters $\beta_1 = 0.5, \beta_2 = 0.999$ with the initial learning rate of $0.0001$, which was decayed by a factor of 10 at the 100${\text{th}}$ epoch. Batch size for training was set to 1. The discriminator was updated once at every update step of the generator. For pre-processing, we normalize the image by the standard deviation of each slice. For $k$-space sub-sampling, we use 1D Gaussian random sampling with the acceleration factor of $R = 3, 4$, selected randomly at each update step. The masks generated for the acceleration factor $R = 3, 4$ also contains central autocalibration signal (ACS) region which take up 6\% and 11\% of the $k$-space central lines, respectively. All experiments were performed using PyTorch. Training took about a day using NVIDIA GeForce GTX 2080-Ti GPU.

\subsection{Comparative Algorithms}

We performed comparative study using the state-of-the-art methods for MR super-resolution and motion artifact correction: ESRGAN~\citep{wang2018esrgan}, cycleGAN~\citep{zhu2017unpaired}, and Oh et al.~\citep{oh2020unsupervised}.

First, we compared our method to ESRGAN, since this is the method utilized in many variants of MR super-resolution \citep{albay2018diffusion, fan2020generative}.
Since ESRGAN is based on supervised training, we generated the training dataset by retrospectively down-sampling the training dataset by the size factor of $\times 2$, and up-sampling the data by bicubic upsampling. Subsequently, the network was trained using the same network architectures described in Section~\ref{sec:net_arch} for fair comparison, using the parameters $\lambda = 0.1, \eta=10$.

CycleGAN~\citep{zhu2017unpaired} is also chosen as one of the comparative algorithms, which is widely used when forward/backward mapping between two arbitrary distributions is needed~\citep{sim2020optimal}. The networks utilized in cycleGAN are set to the same as the ones described in Section~\ref{sec:net_arch}. The only difference is that the vanilla cycleGAN architecture requires two sets of generator/discriminator, which requires double the parameters needed compared to the proposed method.

We finally compare our method to the most recent Oh et al.~\citep{oh2020unsupervised}, which is one of the state-of-the-art algorithms for motion correction, and also the most similar to the proposed method. The network architectures were kept as the same with the proposed method. Parameters in the training/testing phase were all set to the same parameters advised in the original paper.

\begin{table*}[!htb]
    \small
    \centering
    	\resizebox{0.6\textwidth}{!}{
    \begin{tabular}{c|c|c|c|c|c|c}
    \Xhline{2\arrayrulewidth}&&&&&&\\[-1em]
         \thead{Image type} & \thead{Metric} & \thead{Input} & \thead{ESRGAN \\ \citep{wang2018esrgan}} & \thead{cycleGAN \\ \citep{zhu2017unpaired}} & \thead{Oh et al. \\ \citep{oh2020unpaired}} & \thead{\textbf{Proposed}} \\ \hline\hline&&&&&&\\[-1em]
         \multirow{2}{*}{\thead{Brain HCP}} & \thead{PSNR[db]} & 28.33 & 29.56 & 30.03 & 32.21 & \textbf{34.17} \\ \cline{2-7}&&&&&&\\[-1em]
         & \thead{SSIM} & 0.651 & 0.657 & 0.883 & 0.922 & \textbf{0.946} \\ \hline&&&&&&\\[-1em]
         \multirow{2}{*}{\thead{Liver DW-MRI}} & \thead{PSNR[db]} & 35.00 & 35.68 & 30.75 & 38.90 & \textbf{40.04} \\ \cline{2-7}&&&&&&\\[-1em]
         & \thead{SSIM} & 0.763 & 0.802 & 0.780 & 0.938 & \textbf{0.961} \\
    \Xhline{2\arrayrulewidth}
    \end{tabular}
    }
    \caption{Quantitative evaluation of simulation study using Brain HCP data, and liver DW-MRI.}
    \label{tbl:results}
\end{table*}

\subsection{Motion Artifact Simulation}
\label{sec:motion_artifact}

For the performed simulation study with HCP dataset, we simulated the motion artifact, following the procedures in \citep{tamada2020motion}, and the specific parameters used in \citep{oh2020unsupervised}. More specifically, the phase shift $\Phi(\kappa_y)$ in eq.~\eqref{eq:motion} is defined as
\begin{equation}
    \Phi(\kappa_y) = \begin{cases}
    \kappa_y \Delta , \, |\kappa_y| > \kappa_0 \\
    0, \, \text{otherwise},
    \end{cases}
\label{eq:phase_shift}
\end{equation}
where $\Delta$ is the random phase variation sampled as $\Delta \sim \text{Unif}(-37, 37)$ at each phase-encoding line, and the threshold $\kappa_0$ is statically set to $\pi/10$.

Motion artifact simulation for liver DW-MRI data follows \citep{tamada2020motion} for breathing motion. Namely, different from Eq.~\eqref{eq:phase_shift}, the phase shift emerges as a sinusoidal function:
\begin{equation}
    \Phi(\kappa_y) = \begin{cases}
    \kappa_y \Delta \sin (\alpha \kappa_y + \beta) , \, |\kappa_y| > \kappa_0 \\
    0, \, \text{otherwise},
    \end{cases}
\label{eq:phase_shift_periodic}
\end{equation}
where $\alpha \sim \text{Unif}(0.1, 5)$, $\beta \sim \text{Unif}(0, \pi/4)$, and $\Delta \sim \text{Unif}(0, 37)$. The phase shift in Eq.~\eqref{eq:phase_shift_periodic} induces periodic motion artifacts in liver DW-MRI.

\section{Experimental Results}
\label{sec:results}

\subsection{Simulation Study}
For quantitative evaluation of the proposed method, we first perform a simulation study using the HCP dataset, which is illustrated in Fig.~\ref{fig:results_HCP}. We can see both quantitatively and qualitatively the superiority of our method compared to other methods. Namely, ESRGAN is able to boost the resolution of the artifact image as shown in Fig.~\ref{fig:results_HCP}(b), but it also sharpens the motion artifacts along with it. CycleGAN in Fig.~\ref{fig:results_HCP}(c) does remove the artifacts and sharpens the image to some extent, but it also introduces unwanted artifacts, and has low quantitative metric. Oh et al. in Fig.~\ref{fig:results_HCP} efficiently removes the motion artifacts, but produces blurry images due to the degradation. On the other hand, the proposed method in Fig.~\ref{fig:results_HCP}(e) successfuly removes motion artifacts while sharpening the image overall, closely approximating the target image. Quantitative metrics depicting peak signal-to-noise-ratio (PSNR) and structural similarity index (SSIM) in Table~\ref{tbl:results} also show that the results are in fact superior.

Furthermore, we present a simulation study using simulated test set retrospectively acquired from DW-MRI scans, as elaborated in Section~\ref{sec:training_dataset}. Results can be seen in Fig.~\ref{fig:results_DWI_sim} (a-e), where we constantly see that the proposed method is able to closely estimate the label image, devoid of artifacts. Moreover, also in terms of quantitative metrics Table~\ref{tbl:results}, the proposed method outperforms Oh et al.~\citep{oh2020unsupervised} by more than 1 db in PSNR. In contrast, other methods other than the proposed method show their own weaknesses. ESRGAN sharpens the motion artifacts along with the boundaries. CycleGAN makes the overall image sharper, but often breaks down as can be seen in the second row of Fig.~\ref{fig:results_DWI_sim}, and has low PSNR values. Oh et al. is excellent at removing the motion artifacts, but the details are omitted.

\subsection{In Vivo Study}
\label{sec:in-vivo}

Experiments with real DW-MRI data can be seen in Fig.~\ref{fig:results_DWI}. Consistent with the results from the simulation study, we again observe that our method in Fig.~\ref{fig:results_DWI} outperforms all the other methods, effectively removing the motion artifacts while making the image sharper. In contrast, reconstruction results with cycleGAN in Fig.~\ref{fig:results_DWI}(b) does sharpen the image, but introduces unwanted artifacts, and changes the content of the original image, best seen in the third row of the figure. ESRGAN in Fig.~\ref{fig:results_DWI}(c) sharpens the image by a little margin, but the artifact remains intact. Oh et al.~\citep{oh2020unsupervised} in Fig.~\ref{fig:results_DWI}(d) does remove the motion artifacts, but the image still looks blurry.

Moreover, zooming into the images shown in the second row of Fig.~\ref{fig:results_DWI}, we see the region of motion artifact enclosed in the yellow box. We observe that the boundaries are blurry, and thus the structure is not clearly seen in all methods except ESRGAN and the proposed method. However, the proposed method is able to remove the artifacts so that we can visualize the sharp boundaries again. ESRGAN is also able to sharpen the edges, but it also sharpens the artifacts.

\subsection{Quality Enhancement of ADC Maps}
\label{sec:ADC_results}

\begin{figure}[!hbt]
    \centering\includegraphics[width=8cm]{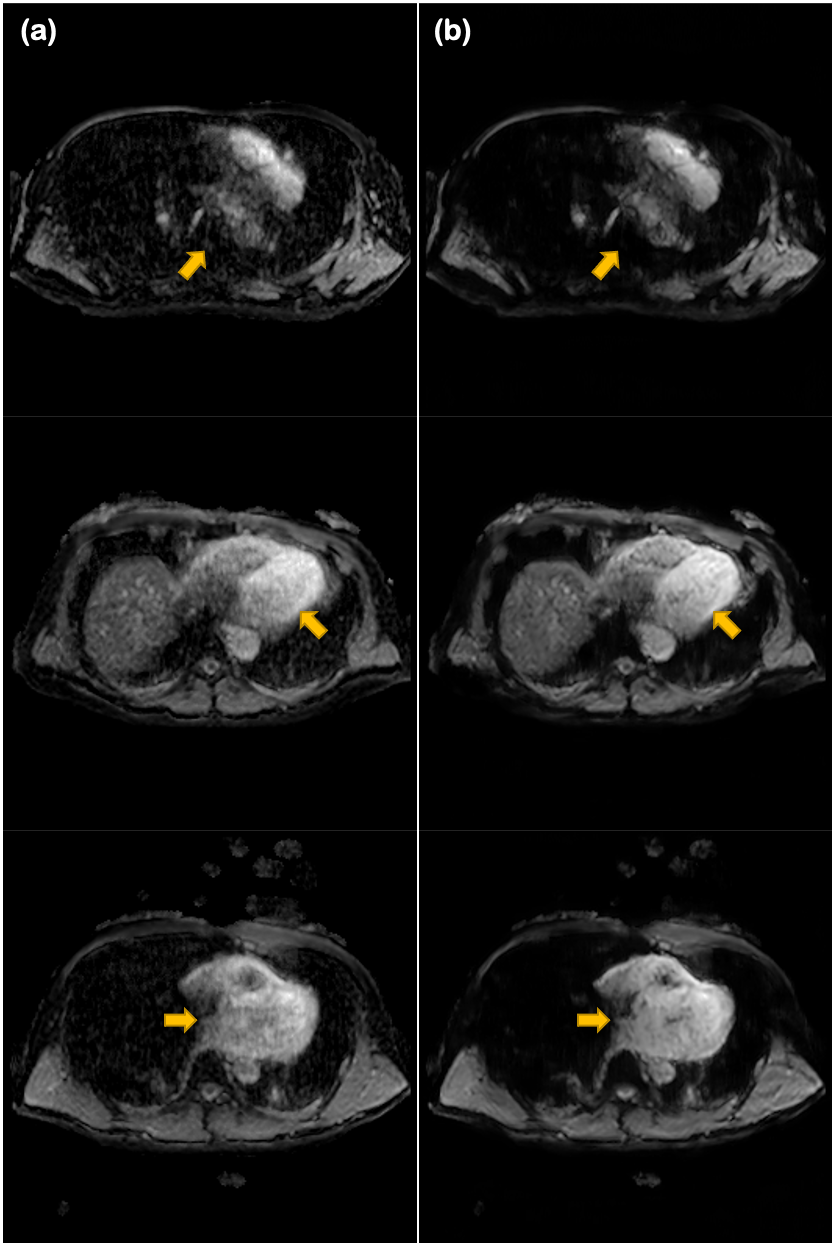}
    \caption{Proposed method directly applied to ADC maps. (a) ADC maps calculated with the conventional method (acquired from vendor) (b) ADC maps reconstructed via the proposed method.}
	\label{fig:ADC_recon}
\end{figure}

The calculation of ADC map via conventional methods often results in low resolution, noisy maps, mostly due to the pixel-wise regression from the different images of varying $b$ values. When the images are not perfectly aligned, this also contributes to the error in the ADC maps, making it harder to estimate the high quality version of the ADC map. We conjectured that since our network learned the ability to enhance the resolution and decrease noise while compensating for motion artifacts, our network could also be applied to ADC maps directly. The results are demonstrated in Fig.~\ref{fig:ADC_recon}. The network used for DW-MRI reconstruction used in Section~\ref{sec:in-vivo} was used to enhance the quality of ADC maps acquired from the vendor using conventional method~\citep{rosenkrantz2011diffusion, koh2007diffusion}. We see in Fig.~\ref{fig:ADC_recon}(a) that the original maps are quite noisy, which could hamper clinical utility. On the other hand, the reconstruction results with the proposed method in Fig.~\ref{fig:ADC_recon} removes noise and enhances the resolution, creating a {\em cleaner} version of the ADC maps.

\begin{figure*}[!hbt]
    \centering\includegraphics[width=18cm]{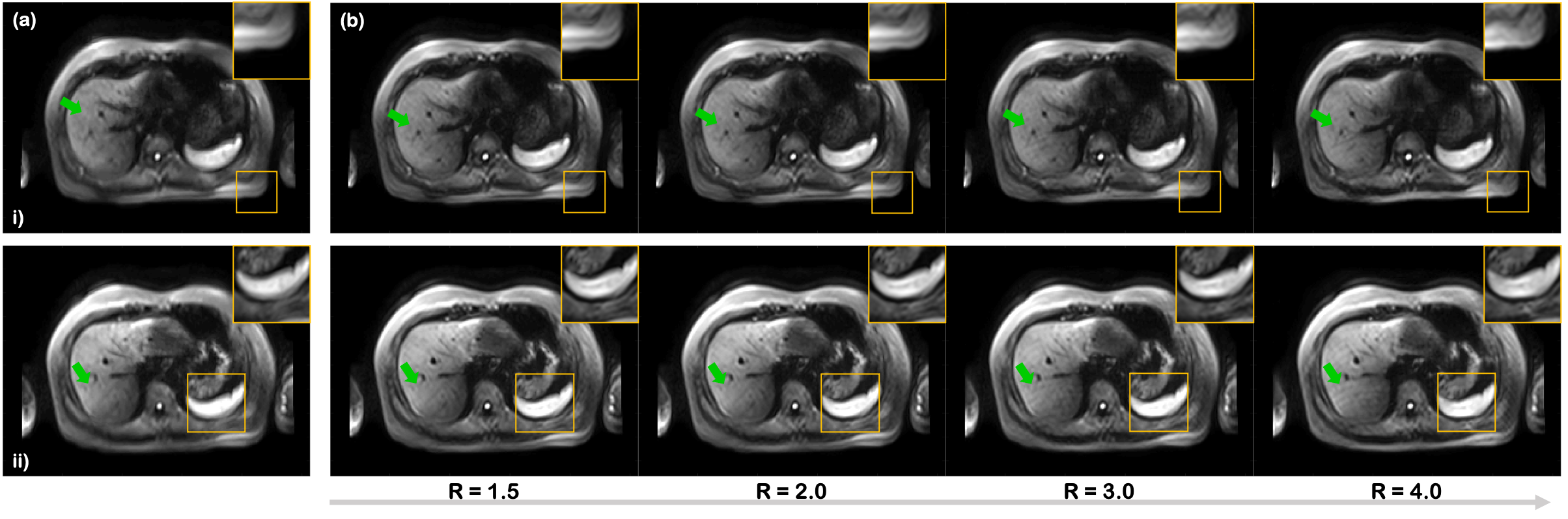}
    \caption{Effect on the reconstructed results by varying the acceleration factor at test stage.  (a) Input artifact images with two different artifact i),ii). (b) Reconstruction results at different acceleration factors. Yellow arrows indicate the area where bulk motion artifacts are observed. Green arrows point where minor aliasing artifacts are introduced.}
	\label{fig:varying_acc_factor}
\end{figure*}

\section{Discussion}
\subsection{Effect of varying sub-sampling factor}
\label{sec:sampling_factor}

It should be noted that the same network can be utilized differently, by changing the bootstrap sub-sampling factor at the inference stage, depicted in Fig.~\ref{fig:net_arch}(b). Specifically, the sub-sampling operator $\Tc$ can be chosen freely, as long as the acceleration factor does not exceed beyond the maximum factor that was used in the training step. In our case, since at the training stage the factor was chosen as $R = 3, 4$, we performed reconstruction using $R = 1.5, 2, 3, 4$, and the results are reported in~\ref{fig:varying_acc_factor}. With both examples, we see that on low acceleration factors, images tend to be cleaner and sharper. As the acceleration factor increases however, we observe that motion artifacts are better corrected, but minor aliasing artifacts are introduced along with it, making the image less clean. Hence, radiologists would be able to control the factor, switching the sub-sampling factor to higher values when more motion correction is desired.

\subsection{Ablation study}

One might question the importance of the degradation block being stochastic. Many factors account for this: for one, the amount of degradation varies across the low quality images, and hence the network needs to learn the variability in the degradation process. Furthermore, sampling the degradation block from certain distribution also has augmentation effect, which could also enhance the expressivity of the network. To strengthen our statement, we performed an ablation study using simulated DW-MRI dataset, as shown in Table~\ref{table:ablation}.

\begin{table}[!hbt]
\begin{center}
	{
    \begin{tabular}{l|c|c}
        \hline
    Type of $\Lc_{\boldsymbol{\psi}}$ & PSNR[db] & SSIM\\
    \hline\hline
    kernel: \xmark, \, noise: \xmark & 35.94 & 0.926 \\
    kernel: \xmark, \, noise: \cmark & 38.68 & 0.959 \\
    kernel: \cmark, \, noise: \xmark & 39.97 & 0.960 \\
    kernel: \cmark, \, noise: \cmark & \textbf{40.04} & \textbf{0.961} \\
    \hline
    \end{tabular}}
\caption{Ablation study on the stochasticity of the degradation block, $\Lc_{\boldsymbol{\psi}}$. For kernel, \xmark: fixed $\sigma_k = 0.5$, \cmark: $\sigma_k \sim \text{Beta}(2.0, 2.0)$. For noise, \xmark: no noise added, \cmark: $\sigma_n = 0.01$.}
\label{table:ablation}
\end{center}
\end{table}

Notably, we can see that keeping the degradation block stochastic, along with injecting random noise produce the best results. Without the noise factor, the metrics slightly decrease, possibly due to noise enhancement in the super-resolution process. Using fixed kernel drops the metrics even more, and finally, when the kernel is fixed and no noise is injected, the performance drops heavily, and we can clearly see the importance of the stochasticity in $\Lc_{\boldsymbol{\psi}}$.

\subsection{Limitations in Acquiring High Quality Target Distribution}

Unsupervised learning, especially in the context of image enhancement, is related to learning the optimal transport mapping from the low-quality input distribution to the high-quality target distribution~\citep{sim2020optimal}. Consequently, a crucial factor which controls the ability of the network is the quality of the training target distribution. When the networks are trained with well-maintained, superior quality images constituting the target distribution, the network learns well, and also performs well at test stage. Unfortunately, the high-quality target distribution utilized in the in vivo study in Section~\ref{sec:in-vivo} share the same scan acquisition parameters with the low-quality input distribution. For example, the number of excitations (NEX) in EPI sequence is the decisive factor of image quality, which is set to 2 in both input and target images.  Although better quality images are manually curated by experienced radiologists for use as a target dataset, this still places limitations on the degree of quality improvement that can be achieved by the trained network. If we were to acquire unpaired higher quality scans with higher NEX as the target distribution, which is in fact possible in DW-MRI, we conjecture that our network will be able to enhance the quality of DW-MRI even further. Having said that, our novel method, which is a proof-of-concept, can further be enhanced in a future study using a well controlled clinical dataset.

\section{Conclusion}
\label{sec:conclusion}

In this work, we proposed a novel unsupervised deep learning method for simultaneous super-resolution and motion artifact correction. The network is trained with the physics-informed OT-cycleGAN which learns to invert the blurring process and reconstruct the under-sampled $k$-space. Subsequently, the trained network can be flexibly used in bootstrap sub-sampling and aggregation process to enhance the low quality DW-MRI images. With extensive experiments using both  simulated and in vivo studies, we showed clear superiority of our method over existing methods. Our framework is flexible in that it can be extended to other imaging situations and modalities, not to mention arbitrary artifacts that are hard to define or solve in a supervised fashion. Applying our method of quality enhancement to other domains could be an interesting directory of study. Finally, the proposed work does not fully analyze the impact and influence of the quality enhancement in clinical perspective, which could also be a promising direction of research.

\section*{Acknowledgments}
This work was supported by the National Research Foundation of Korea under Grant NRF-2020R1A2B5B03001980


\bibliographystyle{model2-names.bst}\biboptions{authoryear}
\bibliography{refs}

\end{document}

%% file: packages.tex
\usepackage{outline}
\usepackage{amssymb}
\usepackage{amsmath}
\usepackage{graphicx}
\usepackage{times}
\usepackage{bm} 
\usepackage{url}

\usepackage{epstopdf}

\usepackage{epsfig}
\usepackage{tikz}
\usetikzlibrary{spy}
\usepackage{algpseudocode}
\usepackage{algorithm}
\usepackage{mathrsfs}

\usepackage{nicefrac}       
\usepackage{booktabs}       

\usepackage{amsmath,graphicx,caption,mathtools,amssymb,amsthm}
\usepackage{graphicx}
\usepackage{multirow}
\usepackage{subcaption}
\usepackage{accents}


%% file: macros.tex
\long\def\comment#1{} 

\newcommand{\xmath}[1] {\ensuremath{#1}\xspace}
\newcommand{\blmath}[1] {\xmath{\bm{#1}}}

\newcommand{\x}{\blmath{x}}

\newcommand{\hb}{{\blmath h}}

\newcommand{\nb}{{\blmath n}}

\newcommand{\xb}{{\blmath x}}
\newcommand{\yb}{{\blmath y}}
\newcommand{\zb}{{\blmath z}}

\newcommand{\Cc}{\mathcal{C}}
\newcommand{\Lc}{\mathcal{L}}
\newcommand{\Nc}{\mathcal{N}}

\newcommand{\Tc}{\mathcal{T}}
\newcommand{\Xc}{\mathcal{X}}

\newcommand{\Zc}{\mathcal{Z}}

\newcommand{\psib}{{\boldsymbol{\psi}}}

\newcommand{\beq}{\begin{equation}}
\newcommand{\eeq}{\end{equation}}
\newcommand{\beqa}{\begin{eqnarray}}
\newcommand{\eeqa}{\end{eqnarray}}
\newcommand{\Fc}{{\mathcal F}}